# Ultra-compact branchless plasmonic interferometers

*Martin Thomaschewski, Yuanqing Yang, Sergey I. Bozhevolnyi*

Centre for Nano Optics, University of Southern Denmark, Campusvej 55, DK-5230 Odense M, Denmark

**ABSTRACT** Miniaturization of functional optical devices and circuits is a key prerequisite for a myriad of applications ranging from biosensing to quantum information processing. This development has considerably been spurred by rapid developments within plasmonics exploiting its unprecedented ability to squeeze light into subwavelength scale. In this study, we investigate on-chip plasmonic systems allowing for synchronous excitation of multiple inputs and examine the interference between two adjacent excited channels. We present a branchless interferometer consisting of two parallel plasmonic waveguides that can be either selectively or coherently excited via ultra-compact antenna couplers. The total coupling efficiency is quantitatively characterized in a systematic manner and shown to exceed 15% for small waveguide separations, with the power distribution between the two waveguides being efficiently and dynamically shaped by adjusting the incident beam position. The presented design principle can readily be extended to other configurations, giving new perspectives for highly dense integrated plasmonic circuitry, optoelectronic devices, and sensing applications.

## Introduction

The past few decades have witnessed impressive progress in the development of miniaturized electronic and photonic devices, driven by the ever-growing demand for faster information transfer and processing capabilities. Plasmonics – the science of utilizing engineered metallic structures to trap, guide and manipulate light via surface plasmon polaritons (SPPs) – has empowered us to break the diffraction limit and pave the way for truly nanoscale optical circuits with unprecedented integration and desired functionality[1–5]. In this regard, plasmonic waveguides, serving as the workhorses of such nanocircuits, have received considerable attention becoming a subject of intensive studies[6–12]. A plethora of various waveguide geometries and device configurations has thus been demonstrated, with fundamental breakthroughs and applications in various areas such as optical communication[12–17], biosensing[18–20] and quantum optics[21–23]. In the quest to realize plasmonic nanocircuits with increasing complexity and compactness, it is indispensable to address the critical issue of crosstalk caused by adjacent transmission channels. Indeed, in any modern integrated circuits with massive parallelism, electromagnetic interference and resulting crosstalk always exist in a general sense and can significantly affect the performance of a system. However, despite tremendous advances in the field of plasmonic nanocircuits, the majority of previous studies have only focused on individual waveguides or serial architectures with a single input[24–28].

Here, we report a theoretical and experimental study of plasmonic nanosystems with closely packed waveguides enabling multiple inputs and parallel communication at telecom wavelength. Just like their electronic and microwave counterparts, plasmonic parallel transmission systems provide the cornerstone for large integrated nanocircuits and promise many practical applications due to their high speed and enhanced information density. Selective and simultaneous excitation of individual or both waveguides in the system are demonstrated by exploiting an ultra-compact nanoantenna coupler. The coupling efficiency of the system is quantitatively characterized in a systematic manner. We show that up to 17% of the free-space incident energy from the air side can be synchronously coupled to two propagating plasmon modes, and that by varying the relative position of the incident beam the power ratio between the two modes can be dynamically tuned. Furthermore, such parallel waveguiding geometry naturally offers itself the ability to function as a fundamental interference unit, without any additional bends or splitters. This enticing property can directly mitigate insertion or bending losses and make the structures substantially easier to fabricate. Developing further this idea, we propose and demonstrate a branchless plasmonic Mach-Zehnder interferometer (MZI), observing its coherent output from the two transmission channels. We envision that our findings can substantially facilitate further miniaturization of many plasmonic devices. For instance, the most compact plasmonic modulators demonstrated so far[13] can substantially be shrunk by replacing bulky grating couplers and splitting elements with the presented parallel design.

## Results and Discussion

The proposed configuration (Fig. 1a) consists of two subwavelength-separated plasmonic slot waveguides, individually fed by impedance-matched nanoantennas[26,29] considered as one of the most compact solutions for a slot waveguide excitation. The pair of dipole antennas with a common side reflector is illuminated with a diffraction-limited focused Gaussian beam polarized along the long axis of the antennas. Electromagnetic radiation is captured by the antenna pair and the induced resonant charge oscillations in each individual antenna are launching surface plasmons polaritons (SPP) into the plasmonic slot waveguides (PSWs). The characterization of the proposed double PSW system is carried out in two steps. In the first step, we use the finite-difference time-domain (FDTD) method[30] to extensively investigate the modal properties supported by this system. The geometrical requirements for a coherent transmission with low cross-talk are defined. On this basis, the structures are designed, fabricated and characterized.

We start by examining the properties of a single plasmonic slot waveguide that constitute the transmission system in our study. To reduce mode leakage into the glass substrate ($n_{glass}$ = 1.45), a 300-nm-thick dielectric layer (PMMA, $n_{PMMA}$ = 1.49) is used to cover the PSW. Here, we focus on the PSW modes at the telecom wavelength $\lambda_0 = 1550$ nm, and employ the optical constants of gold reported by Johnson and Christy[31]. A representative field distribution of the fundamental mode supported by single PSWs is shown in Fig. 2a. The function of its modal effective index and propagation length with respect to slot width $W_{slot}$ and variant metal thickness $H$ are depicted in Fig. 2c. With decreasing slot width, a larger overlap between the mode profile and the metal gives rise to an increase in the model index accompanied with a decrease in the propagation length. To maintain a relatively low loss characteristic while keeping the whole device thin and compact enough, we choose a balanced set of geometric sizes as $H = 100$ nm and $W_{slot} = 300$ nm for the single PSWs, as indicated by the star symbol in Fig. 2c. Then we investigate the optical transmission system consisting of two parallel PSWs, which is in the following stated as the double PSW. Different from the single PSW, the double PSW system can support two orthogonal eigenmodes, characterized by anti-symmetric $E_a(x,y)$ and symmetric $E_s(x,y)$ field distributions, as shown in Fig. 2b. The total electric field in the system $E_{total}$ can thus be expressed as a linear superposition of the two modes: $E_{total} = A_a(z)E_a + A_s(z)E_s$, where modal amplitudes $A_a(z)$, $A_s(z)$ can be given by $A_a(z) = A_a(0)\exp(-j2\pi n_a z/\lambda_0)$ and $A_s(z) = A_s(0)\exp(-j2\pi n_s z/\lambda_0)$ with the corresponding effective modal indices $n_a$ and $n_s$ of the anti-symmetric and symmetric modes, respectively. In Fig. 2d, we plot the functions of these two modal indices $n_a$ and $n_s$ with respect to waveguide separation $W_{sep}$. An asymptotic behaviour can be clearly seen when $W_{sep}$ approaches the limit $W_{sep} \sim \infty$. In this case, both eigenmodes approximate the fundamental mode of the single PSW and thus the whole system function as two uncoupled PSWs (Supplementary Fig. S1). By contrast, when $W_{sep}$ gradually approaches to the other limit $W_{sep} \sim 0$, a large spatial overlap between the fields inside two closely adjacent PSWs would cause significant coupling, which can be quantified by the beating length $L_c = \lambda_0/2|n_a - n_s|$. A smaller $L_c$ implies stronger coupling and associated higher crosstalk between the two channels. To further unambiguously reveal the influence of the crosstalk in practical devices, in Fig. 2e we show the crosstalk between the two channels for different device lengths $L$. We evaluate the crosstalk $C$ by defining it as $C = P_t/P_0$, where $P_t$ is the power transferred from one waveguide to the other and $P_0$ is the power of the original signal. With the mean propagation length $L_p = \lambda_0/2\text{Im}(n_a - n_s)$ and an assumption that $\text{Im}(n_a - n_s) \ll \text{Re}(n_a - n_s)$, the crosstalk can be calculated as follows[32]:

$$C \cong 10\log_{10}\left[\exp\left(-2\frac{L}{L_p}\right)\sin^2\left(\frac{\pi}{2}\frac{L}{L_c}\right)\right].$$

The sine function in the above equation represents the periodic feature of the coupling event occurring between the two PSWs, which can also be easily identified from Fig. 2e while the first exponential term renders the loss factor of the coupling. To avoid the fluctuations for different device lengths and minimize the unfavourable crosstalk (< −10 dB), we focus in the following experimental study on the design where the two PSWs are only in weak-coupling regions, as denoted by the shaded grey area in Fig. 2d.

For the experiment, all investigated structures are fabricated on $SiO_2$ using electron-beam lithography (EBL) followed by thermal gold evaporation and subsequent metal lift-off. A 300-nm thin layer of PMMA is deposited by using the spin-coating technique. The scanning electron microscope (SEM) image in Fig. 1c depicts a fabricated double PSW with the zoomed view of one nanoantenna pair. Transmission measurements were performed with a home-built inverted microscope setup, having the sample mounted on a three-dimensional stabilized piezo translation stage with 20 nm spatial resolution. A linearly polarized laser beam with the wavelength of $\lambda_0 =$

1550 nm was focused by a high numerical aperture objective (NA = 0.95 dry) to a diffraction-limited FWHM width of $\emptyset = (2.3 \pm 0.1)$ µm. We first demonstrate the transmission experiment on a structure with a waveguide separation of 1.1 µm, to ensure that the two waveguides can be selectively excited while the scattering signal of each individual antenna can still be resolved. The excitation of the two parallel gap modes was shown to be sensitive to the relative position between the feeding point (i.e. center of incident Gaussian beam) and antenna pair. For comparison, Fig. 3a-c shows the reflection images captured by the camera with three different relative beam positions, with the output spot moving in accord with the incident beam positioning from the upper out-coupling antenna (Fig. 3a) to the lower one (Fig. 3c). For the symmetric location of an incident laser beam, both waveguides are excited equally with two output spots indicating coherently excited and spatially separated wavefronts transmitted from the antennas (Fig. 3b). It is thereby shown that the adjustment of the focused incident beam location on the antenna pair structure enables efficient shaping of the power distribution routed through the double slot waveguide system.

To investigate quantitatively the in-coupling conditions, we designed several structures with varying waveguide separation distances. The total coupling efficiency $C_{tot}$ can be determined by measuring the incident optical power $P_{in}$ and the optical power $P_{out}$ radiated from the transmitting antenna pair. The incident power $P_{in}$ is estimated by measuring the power $P_R$ reflected from a flat gold film on the investigated sample, and by assuming $P_{in} = P_R \cdot T_{PMMA}^2 \cdot R_{Au}$ with the transmission coefficient $T_{PMMA} = 0.96$ of 300 nm thick PMMA and the reflectance coefficient $R_{Au} = 0.98$ of 100 nm Au for $\lambda_0 = 1550$ nm. Due to reciprocity and symmetry of the system, the coupling characteristics of the receiving antenna pair are identical with the antenna pair of transmission. Considering an exponential intensity decay of the electric field amplitude along the propagation direction the total coupling efficiency can be expressed as $C_{tot} = C_{Ant} \cdot e^{-L/L_p} \cdot C_{Ant} = C_{Ant}^2 \cdot e^{-L/L_p} = P_{out}/P_{in}$, where $L_p$ is the propagation length of the two waveguide modes and $C_{Ant}$ is the coupling efficiency of the antenna pair. To calculate the coupling efficiency $C_{Ant}$, we first determine the propagation length $L_p$ by measuring the transmission for a series of double PSW systems with different lengths and the excitation beam centered at the input antenna pair (See Fig. 3d). The calculated propagation length of 12.3 µm is in good agreement with the measured propagation length of 12.0 µm, which is used in the following for the calculation of the in-coupling efficiency of the antenna pair. Figure 4a depicts the measured in-coupling efficiencies exemplarily for four different waveguide separations as a function of the relative beam position and the waveguide separation. Structures with a large waveguide separation distance above 800 nm show a maximized coupling efficiency, when the incident beam coincides with one of the individual antennas. At this regime of separation, the energy provided by the diffraction-limited beam is mostly directed into one waveguide. For smaller waveguide separations, we observed a maximized coupling efficiency when the beam is positioned between the individual antennas. Here, more power is captured into the double waveguide system, due to a closer match between excitation spot and effective area of the antenna pair. It is noteworthy that, although the excitation beam is highly focused, and the maximum of the Gaussian beam is positioned between the antennas, it can still exceed the efficiency obtained by a single waveguide system by 10 % (See Fig. 4b).

On-chip interference can be demonstrated by branching the two simultaneous excited channels. Therefore, we fabricated a set of MZI with different physical path lengths between both interferometer arms (Supplementary Fig. S2), and thus inducing a phase difference between the two channels ranging from 0 and $2\pi$. As a result, the optical signal collected from the single antenna at the output port is modulated by constructive or destructive interference (Fig. 5). For further miniaturization of the interferometer, we investigated branchless systems with waveguide separation below 800 nm. At this regime of separation, we noticed an overlap of the signal emitted from the two out-coupling antennas driven by the transmitted signal. The two lobes in the image plane merge due to spatially overlapping wave fronts. Due to the far-field interference of the waveguide-driven nanoantennas, the scattering spot can reveal information about the relative phase and intensity of the individual emitters[33,34]. This finding implies that the waveguide system acts as an ultra-compact and branchless Mach-Zehnder interferometer. To investigate this phenomenon, we fabricated an asymmetric system consisting of a waveguide with 150 nm and 300 nm slot width. The difference $\Delta n$ in the mode effective index introduces a relative modal phase shift of $\Delta\varphi = \Delta n \cdot L$, where $L$ is the length of the waveguide system. The simulated phase evolution mapped in Fig. 5a and 5b for a 15.63 µm and a 7.81 µm long structure shows a $2\pi$ and $\pi$ phase difference, respectively. We illustrate perfect constructive and destructive by measuring the transmission of the individual systems. Although the constructive interfering system is much longer, it shows a much higher transmission than the shorter system, where the signal emitted from the antennas are overlapping and destructively interfering in the far-field. It is noteworthy, that unequal losses in the individual waveguides and differences in coupling efficiencies can potentially be compensated by adjusting the beam position. Here, the maximum of the Gaussian beam is still positioned nearly the center of the in-coupling nanoantenna pair but the transmitted intensity is maximized for the in-phase or minimized for the out-of-phase configuration, respectively, by slightly shifting the relative beam position from the antenna pairs centroid point. Based on this capability, a high extinction ratio of 18.3 dB is observed. This value is in good agreement with the expected extinction ration of 20 dB estimated by calculating the far-field interference pattern produced by two separated point-like dipole emitters under consideration of the unequal structure lengths and the finite acceptance angle of the used microscope objective. As shown in Fig. 5c, the resulting far-field interference of the in-phase radiating antennas yield in an intensity distribution mainly directed toward the normal direction. However, assuming two dipoles oscillating out-of-phase, the wavefronts cancel towards the normal direction due to destructive inference, while showing the first-order maxima emitting at large angles, barely detectable by the used high NA objective.

## Conclusion

Summarizing, we have proposed and investigated, theoretically and experimentally, a branchless waveguide configuration that allows the simultaneous coherent excitation of of two plasmonic slot waveguide modes. The subwavelength separated parallel slot waveguides fed by ultra-compact nanoantennas show a reasonably large in-coupling efficiency without significant cross-talk between the two individual slot waveguides. This system is utilized to realize a novel ultra-compact Mach-Zehnder interferometer without a requirement for on-chip waveguide branching, making this system to be a very compact solution for sensing applications. Besides the miniaturization capabilities, another advantage of our design is the lateral accessibility of the metal pads forming this interferometric system. This unique attribute makes it a promising candidate for integrated plasmonic electro-optic applications.

**Conflicts of interest**

There are no conflicts of interest to declare

**Acknowledgements**

This work was funded by the European Research Council (the PLAQNAP project, Grant 341054) and the University of Southern Denmark (SDU2020 funding).

**References**

1  E. Ozbay, *Science*, 2006, 311, 189–193.

2  S. A. Maier, *Plasmonics: Fundamentals and Applications*, Springer, 2007.

3  D. K. Gramotnev and S. I. Bozhevolnyi, *Nat. Photonics*, 2010, **4**, 83–91.

4  T. J. Davis, D. E. Gómez and A. Roberts, *Nanophotonics,* 2016, **6**, 543–559.

5  Z. Fang and X. Zhu, *Adv. Mater.*, 2013, **25**, 3840–3856.

6  Y. Fang and M. Sun, *Light-Sci. Appl.*, 2015, **4**, e294.

7  Z. Han and S. I. Bozhevolnyi, *Rep. Prog. Phy.s*, 2013, **76**, 016402.

8  M. Großmann, M. Thomaschewski, A. Klick, A. Goszczak, E. Sobolewska, T. Leißner, J. Adam, J.Fiutowski, H.-G. Rubahn and M. Bauer, *Plasmonics*, 2017, **1–8**.

9  W. Cai, W. Shin, S. Fan and M. L. Brongersma, *Adv. Mater.*, 2010, **22**, 5120–5124.

10  C. L. Smith, N. Stenger, A. Kristensen, A. N. Mortensen and S. I. Bozhevolnyi, *Nanoscale*, 2015, **7**, 9355–9386.

11  M. Cohen, Z. Zalevsky and R. Shavit, *Nanoscale*, 2013, **5**, 5442–5449.

12  C. Haffner, W. Heni, Y. Fedoryshyn, J. Niegemann, A. Melikyan, D. Elder, B. Baeuerle, Y. Salamin, A. Josten, U. Koch, C. Hoessbacher, F. Ducry, L. Juchli, A. Emboras, D. Hillerkuss, M. Kohl, L. Dalton, C. Hafner and J. Leuthold, *Nat. Photonics*, 2015, **9**, 525–528.

13  M. Ayata, Y. Fedoryshyn, W. Heni, B. Baeuerle, A. Josten, M. Zahner, U. Koch, Y. Salamin, C. Hoessbacher, C. Haffner, D. L. Elder, L. R. Dalton and J. Leuthold, *Science*, 2017, **358**, 630–632.

14  D. Ansell, I. Radko, Z. Han, F. Rodriguez, S. Bozhevolnyi and A. Grigorenko, *Nat. Commun.*, 2015, **6**, 8846.

15  Y. Ding, X. Guan, X. Zhu, H. Hu, S. Bozhevolnyi, L. Oxenløwe, K. Jin, N. Mortensen and S. Xiao, *Nanoscale*, 2017, **9**, 15576–15581.

16  J. Chen, C. Sun, H. Li and Q. Gong, *Nanoscale*, 2014, **6**, 13487–13493.

17  Y. Yang, Q. Li and M. Qiu, *Sci. Rep.*, 2016, **6**, 19490.

18  S. Lal, S. Link and N. J. Halas, *Nat. Photonics*, 2007, **1**, 641–648.

19  J. Homola, *Chem. Rev.*, 2008, **108**, 462–493.

20  J. N. Anker, P. W. Hall, O. Lyandres, N. C. Shah, J. Zhao and R. P. Duyne, *Nat. Mater.*, 2008, **7**, 442–453.

21  S. Kumar, A. Huck and U. L. Andersen, Nano Lett., 2013, 13, 1221–5.

22  X. Wu, P. Jiang, G. Razinskas, Y. Huo, H. Zhang, M. Kamp, A. Rastelli, O. G. Schmidt, B. Hecht, K. Lindfors and M. Lippitz, *Nano Lett.*, 2017, **17**, 4291–4296.

23  R. W. Heeres, L. P. Kouwenhoven and V. Zwiller, *Nat. Nanotechnol.*, 2013, **8**, 719–722.

24  K. C. Huang, M.-K. Seo, T. Sarmiento, Y. Huo, J. S. Harris and M. L. Brongersma, *Nat. Photonics*, 2014, **8**, nphoton.2014.2.

25  C. Rewitz, G. Razinskas, P. Geisler, E. Krauss, S. Goetz, M. Pawłowska, B. Hecht and T. Brixner, *Phys. Rev. Appl.*, 2014, **1**, 014007.

26  A. Kriesch, S. P. Burgos, D. Ploss, H. Pfeifer, H. A. Atwater and U. Peschel, *Nano Lett.*, 2013, **13**, 4539–45.

27  S. P. Burgos, H. W. Lee, E. Feigenbaum, R. M. Briggs and H. A. Atwater, *Nano Lett.*, 2014, **14**, 3284–3292.

28  H. W. Lee, G. Papadakis, S. P. Burgos, K. Chander, A. Kriesch, R. Pala, U. Peschel and H. A. Atwater, *Nano Lett.*, 2014, **14**, 6463–6468.

29  A. Andryieuski, R. Malureanu, G. Biagi, T. Holmgaard and A. Lavrinenko, *Opt. Lett.*, 2012, **37**, 1124.

30  Lumerical Inc. http://www.lumerical.com/tcad-products/fdtd/

31  P. Johnson and R. Christy, *Phys. Rev. B*, 1972, **6**, 4370–4379.

32  G. Veronis and S. Fan, *Opt. Express*, 2008, 16, 2129–40.

33  D. Wolf, T. Schumacher and M. Lippitz, *Nat. Commun.*, 2016, **7**, 10361.

34  C. Ropp, Z. Cummins, S. Nah, J. T. Fourkas, B. Shapiro and E. Waks, *Nat. Commun.*, 2015, **6**, 6558.

*Figure 1*

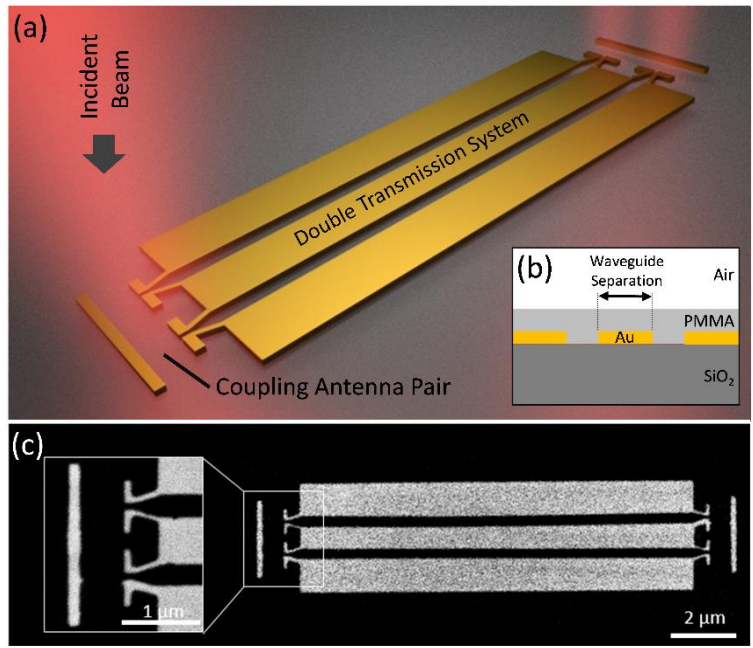

**Fig. 1** *(a) Three-dimensional rendering of the proposed parallel PSWs fed by a pair of optical-loaded nanoantennas with side reflectors at the input and output ports. (b) Schematic cross-section and (c) top-view SEM images of a fabricated device are illustrated.*

*Figure 2*

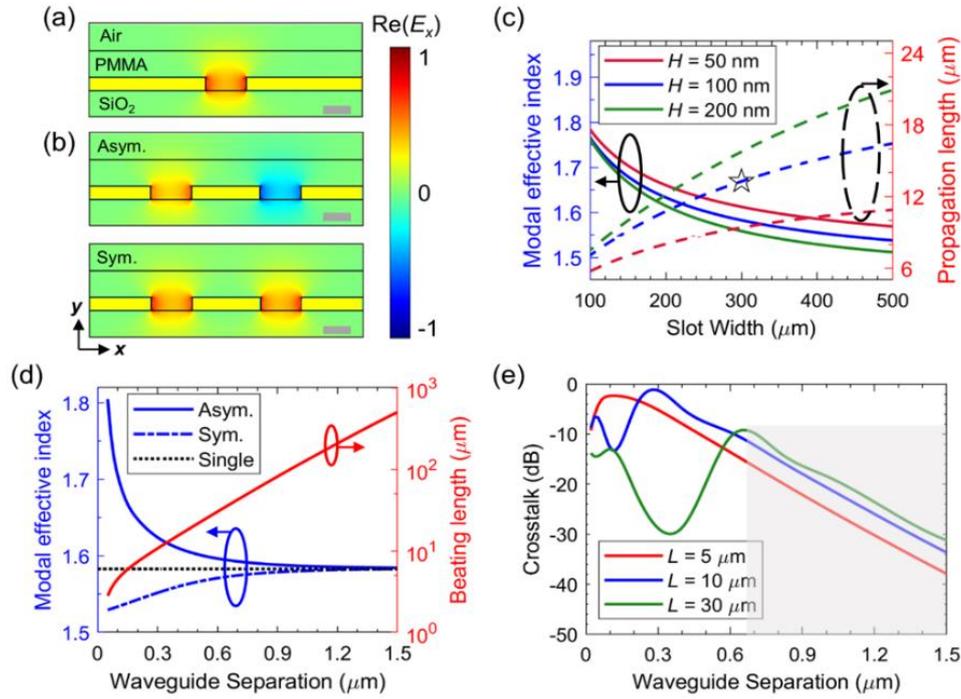

*Fig.2 Analysis of transmission systems consisting of individual or parallel plasmonic slot waveguides. (a) Mode profile Re(Ex) of a single plasmonic slot waveguide with metal thickness H = 100 nm and slot width W = 300 nm. The yellow colour represents the gold material. (b) Field distributions of anti-symmetric (upper panel) and symmetric (lower panel) modes of a parallel plasmonic waveguiding system with H = 100 nm, Wslot = 300 nm and Wsep = 500 nm. The scale bars in (a) and (b) represent 200 nm. (c) Modal effective index (solid lines, left y-axis) and propagation length (dashed lines, right y-axis) of the fundamental mode supported by single plasmonic slot waveguides with variant geometric sizes. The star symbol points out the geometric size used in our study, i.e., H = 100 nm, Wslot = 300 nm. (d) Modal effective index (left axis) of the two eigenmodes supported by the parallel plasmonic waveguiding system with varying separations Wsep and fixed H = 100 nm and Wslot = 300 nm. The black dotted line shows the property of a referenced single plasmonic waveguide with the same H and W. Beating length Lc of the system (red line) is plotted in the right y-axis. (e) Function of the crosstalk between the two parallel plasmonic slot waveguides with respect to separation distance. Scenarios of three different device lengths L are considered and the shaded area indicates the weak-coupling region where the crosstalk C < − 10 dB.*

*Figure 3*

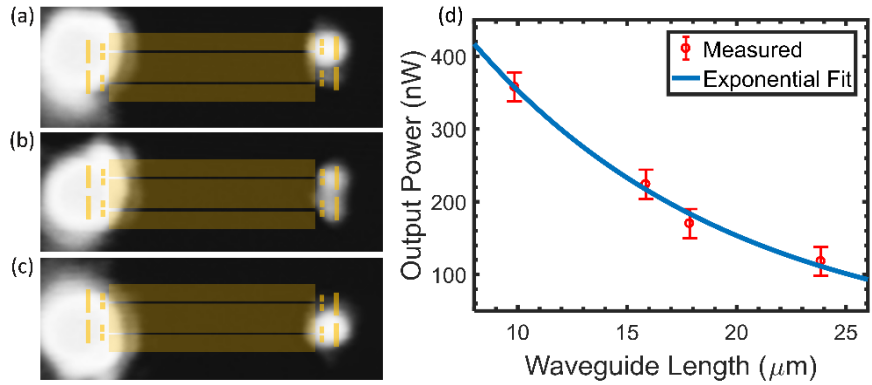

*Fig.3 (a-c) Far-field optical characterization at λ0 = 1550 nm reveals selective waveguide excitation at beam position: (a) beam positioned at the upper antenna, (b) beam positioned centred between nanoantennas and (c) beam positioned at the upper antenna for a double PSWs with 1100 nm separation. The input and output antenna pairs with the waveguide structure are superimposed as a guide to the eyes. (d) Output intensities versus waveguide length are plotted and fitted as a first-order exponential decay function. The blue points represent the measured output power and the red line is the fitting graph.*

*Figure 4*

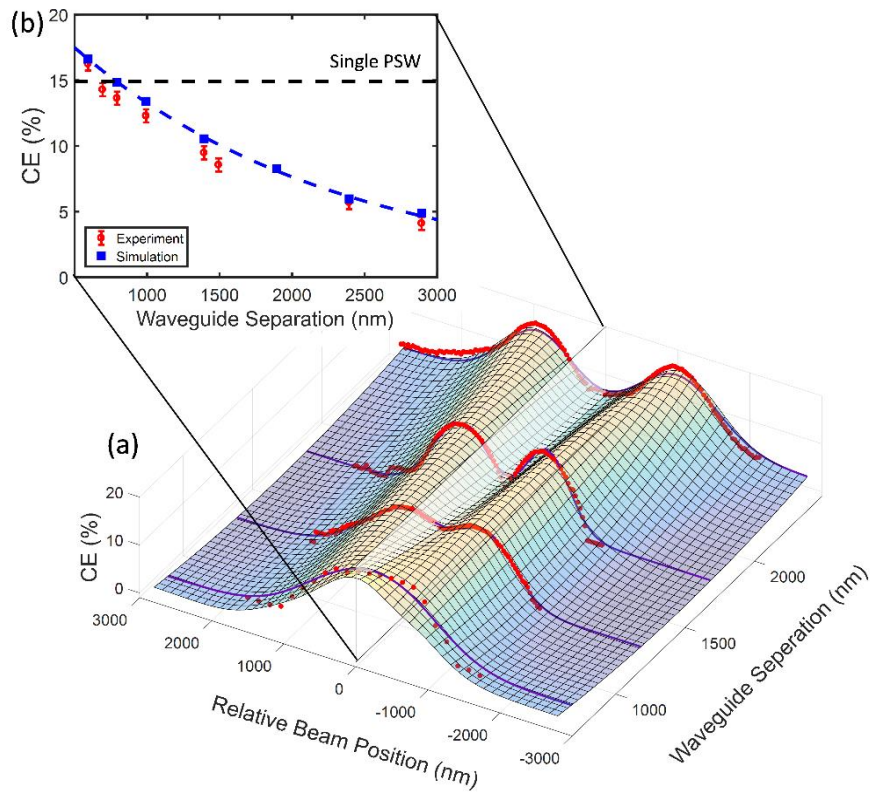

***Fig.4*** *Dependence of in-coupling efficiency on the waveguide separation distance and the relative beam position. (a) Measured coupling efficiencies are exemplarily shown for four different waveguide separation distances (red dots), which dependency on the relative beam position can be well approximated by a sum of two Gaussians, shown as purple lines. (b) Coupling efficiency when the relative beam position coinciding with the centre of the antenna pair, illustrated as a slice in (a). The calculated results are plotted in blue, while the red circles show the efficiencies obtained from the experiment. The black dashed line represents the maximum coupling efficiency measured for a single plasmonic slot waveguide system fed by one single nanoantenna.*

*Figure 5*

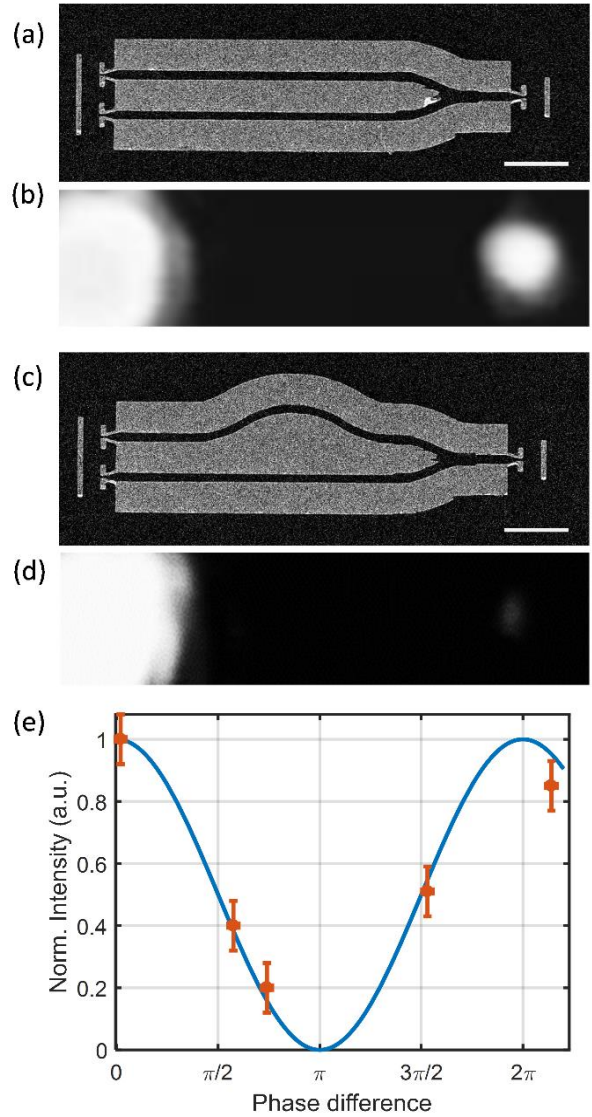

*Fig. 5* SEM image of Mach-Zehnder Interferometer with a nominal optical phase difference of (a) 0 and (c) 3π/4, with the corresponding far-field images (b) and (d). The scale bars in (a) and (c) represent 2 µm. (e) Measured and calculated normalized optical power output versus measured relative phase difference with the beam positioned between the antennas.

*Figure 6*

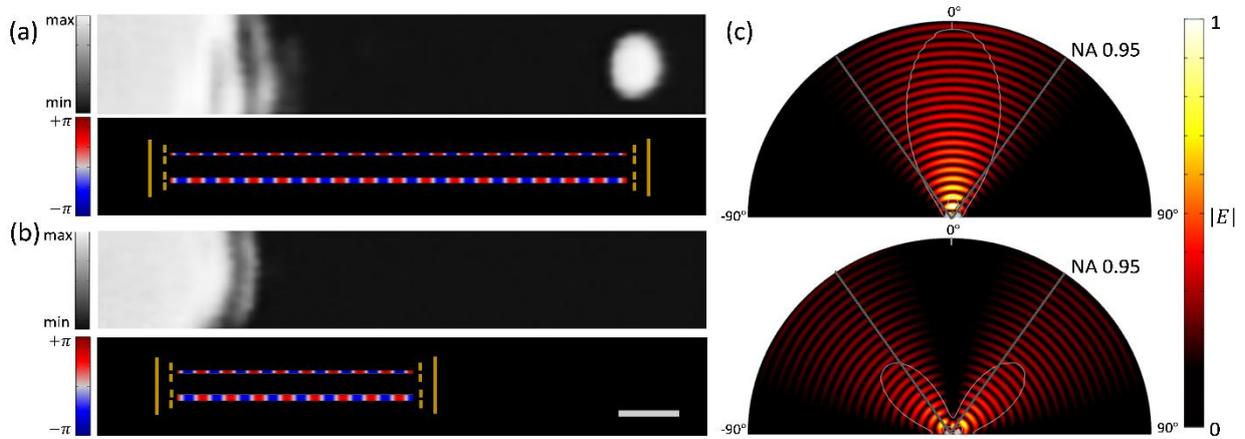

**Fig. 6** *Asymmetric double plasmonic slot waveguides consisting of two waveguides with the width of 150 nm and 300 nm, respectively. By varying the device length, the relative phase is adjusted to be in-phase (a) and out of phase (b). The far-field images (top) and the simulated waves (bottom) and are respectively shown for structures with a waveguide separation of 700 nm. The scale bar represent 1 µm. (c) Simulated far-field interference pattern of two separated dipole emitters radiating in phase (top) and out of phase (bottom). The white line illustrated the radiation pattern and the grey line shows the acceptable detection angle of the used optical system.*